\documentclass[article]{aa}  

\usepackage{graphicx}
\usepackage{txfonts}
\usepackage{lscape}
\usepackage{booktabs}
\usepackage{threeparttablex}
\usepackage{wasysym}
\usepackage{arydshln}
\usepackage{mathrsfs}
\usepackage{amssymb}
\usepackage[bookmarks=false,colorlinks=true,linkcolor=black,citecolor=cyan,filecolor=black,urlcolor=cyan]{hyperref}
\usepackage{array}

\newcolumntype{C}[1]{>{\centering\arraybackslash}p{#1}}

\newcommand{\pfrac}[2]{\left( \frac{#1}{#2} \right)}

\begin{document}

\title{Ticking away: the long-term X-ray timing and spectral evolution of eRO-QPE2}

\author{R. Arcodia\thanks{NASA Einstein fellow}
		\inst{1},
        I. Linial
		\inst{2,3},
        G. Miniutti
		\inst{4},
        A. Franchini
		\inst{5,6},
        M. Giustini
		\inst{4},
        M. Bonetti
		\inst{6,7},
        A. Sesana
		\inst{6,7,11},
        R. Soria
		\inst{8,9},
        J. Chakraborty
		\inst{1},
        M. Dotti
		\inst{6,7},
        E. Kara
		\inst{1}, 
        A. Merloni
		\inst{10}, 
        G. Ponti
		\inst{11,10}
        F. Vincentelli
		\inst{12}
	    }
\institute{MIT Kavli Institute for Astrophysics and Space Research, 70 Vassar Street, Cambridge, MA 02139, USA\\
\email{rarcodia@mit.edu}
    \and
Department of Physics and Columbia Astrophysics Laboratory, Columbia University, New York, NY 10027, USA
    \and
Institute for Advanced Study, 1 Einstein Drive, Princeton, NJ 08540, USA
    \and
    Centro de Astrobiolog\'ia (CAB), CSIC-INTA, Camino Bajo del Castillo s/n, 28692 Villanueva de la Ca\~nada, Madrid, Spain 
    \and
    Universität Zürich, Institut für Astrophysik, Winterthurerstrasse 190, CH-8057 Zürich, Switzerland
    \and
    Universit\`a degli Studi di Milano-Bicocca, Piazza della Scienza 3, I-20126 Milano, Italy
    \and
    INFN, Sezione di Milano-Bicocca, Piazza della Scienza 3, I-20126 Milano, Italy
    \and
   INAF-Osservatorio Astrofisico di Torino, Strada Osservatorio 20, I-10025 Pino Torinese, Italy
   \and
    Sydney Institute for Astronomy, School of Physics A28, The University of Sydney, Sydney, NSW 2006, Australia
   \and
   {Max-Planck-Institut f\"ur extraterrestrische Physik, Giessenbachstrasse, 85748, Garching, Germany} 
   \and
   {INAF -- Osservatorio Astronomico di Brera, Via E. Bianchi 46, 23807 Merate, Italy} 
   \and
{School of Physics and Astronomy, University of Southampton, University Road, Southampton, SO17 1BJ, UK}    }
\date{Received ; accepted }


\abstract{Quasi-periodic eruptions (QPEs) are repeated X-ray flares from galactic nuclei that recur every few hours to days, depending on the source. Despite some diversity in the recurrence and amplitude of eruptions, their striking regularity has motivated theorists to associate QPEs with orbital systems. 
Among the known QPE sources, eRO-QPE2 has shown the most regular flare timing and luminosity since its discovery. We report here on its long-term evolution over $3.3\,$yr from discovery and find that: i) the average QPE recurrence time per epoch has decreased over time, albeit not at a uniform rate; ii) the distinct alternation between consecutive long and short recurrence times found at discovery has not been significant since; iii) the spectral properties, namely flux and temperature of both eruptions and quiescence components, have remained remarkably consistent within uncertainties. 
We attempted to interpret these results as orbital period and eccentricity decay coupled with orbital and disk precession. However, since gaps between observations are too long, we are not able to distinguish between an evolution dominated by just a decreasing trend, or by large modulations (e.g. due to the precession frequencies at play). In the former case, the observed period decrease is roughly consistent with that of a star losing orbital energy due to hydrodynamic gas drag from disk collisions, although the related eccentricity decay is too fast and additional modulations have to contribute too. In the latter case, no conclusive remarks are possible on the orbital evolution and the nature of the orbiter due to the many effects at play. However, these two cases come with distinctive predictions for future X-ray data: in the former, we expect all future observations to show a shorter recurrence time than the latest epoch, while in the latter we expect some future observations to be found with a larger recurrence, hence an apparent temporary period increase.}



	
	\titlerunning{temp title}
	\authorrunning{R. Arcodia et al.} 
	\maketitle	

\section{Introduction}
\label{sec:intro}

Quasi-periodic eruptions (QPEs) are repeating soft X-ray bursts originating from galactic nuclei which last and recur on the timescale of hours \citep{Miniutti+2019:qpe1,Giustini+2020:qpe2,Chakraborty+2021:qpe5cand,Arcodia+2021:eroqpes,Arcodia+2024:qpe34}. Their unique characteristic is the soft X-ray spectrum during the eruptions, and its evolution with a harder rise than decay \citep{Arcodia+2022:ero1_timing,Miniutti+2023:gsnrebr,Arcodia+2024:qpe34}. This peculiar behavior helps distinguishing their physical emission mechanism from that of other repeated nuclear transients, which are growing in numbers and flavors. To date, most models for the origin of QPEs relate to galactic nuclei stellar dynamics, although this is not the sole interpretation \citep[e.g.,][for a recent overview of the models]{Arcodia+2024:qpe34}. In particular, a popular scenario \citep[e.g.,][]{Linial+2023:qpemodel2,Franchini+2023:qpemodel,Tagawa+2023:qpemodel,Zhou+2024:emri,Zhou+2024arXiv240506429Z} that we aim to test in this work involves a primary massive black hole with its compact and short-lived accretion flow \citep[e.g.,][]{Patra2024MNRAS.530.5120P}, which in some cases may be fed by a tidal disruption event \citep[][and references therein]{Linial+2023:qpemodel2,Franchini+2023:qpemodel}; this recent accretion event reveals a separate, pre-existing, much smaller object (e.g., a star or a black hole) on a low-eccentricity orbit, repeatedly ploughing through the disk twice per orbit and producing the observed quasi-periodic eruptions. This interpretation has drawn particular interest since these systems involve the so-called extreme mass ratio inspirals \citep[EMRIs; e.g.,][]{Amaro-Seoane+2018:emris} which (depending on the nature of the orbiter), would be detectable by future-generation gravitational-wave detectors such as the Laser Interferometer Space Antenna (LISA) and $\mu$Ares \citep{2021ExA....51.1333S,2024arXiv240207571C}. If this were to be the correct model, QPE volumetric rates would be the first ever observational constraint for EMRIs \citep{Arcodia+2024:rates}. 

\begin{figure*}[tb]
		\centering
		\includegraphics[width=0.88\textwidth]{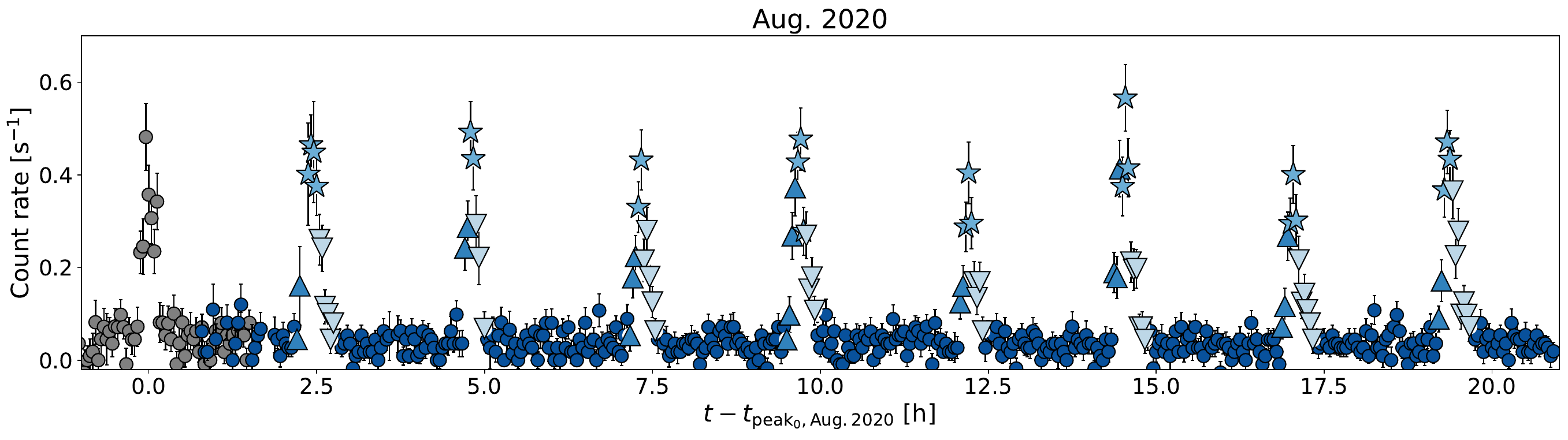}
		\includegraphics[width=0.88\textwidth]{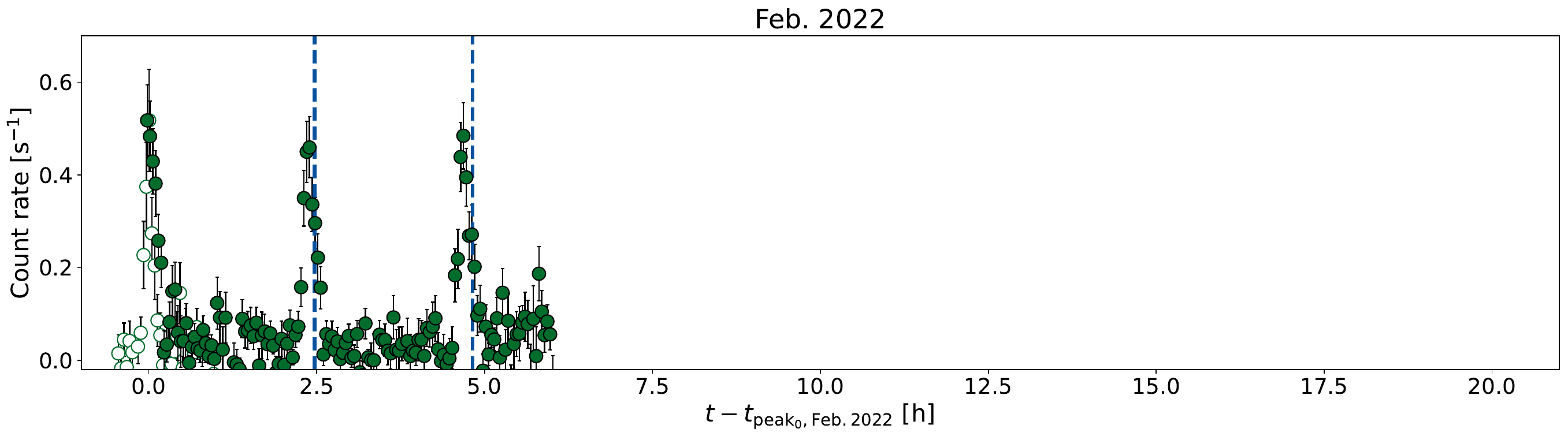}
		\includegraphics[width=0.88\textwidth]{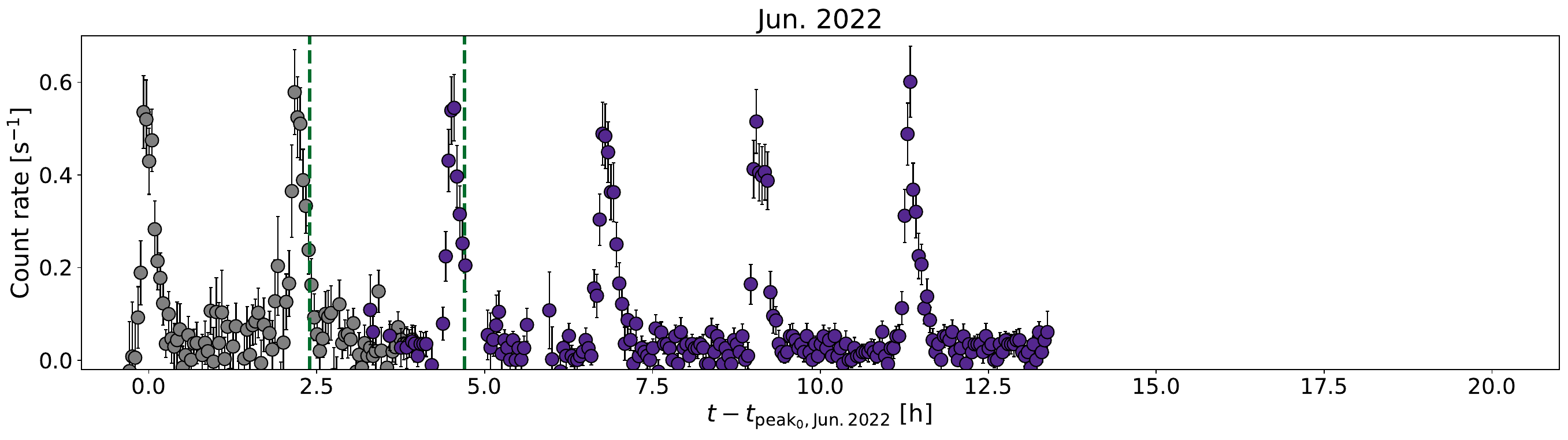}
		\includegraphics[width=0.88\textwidth]{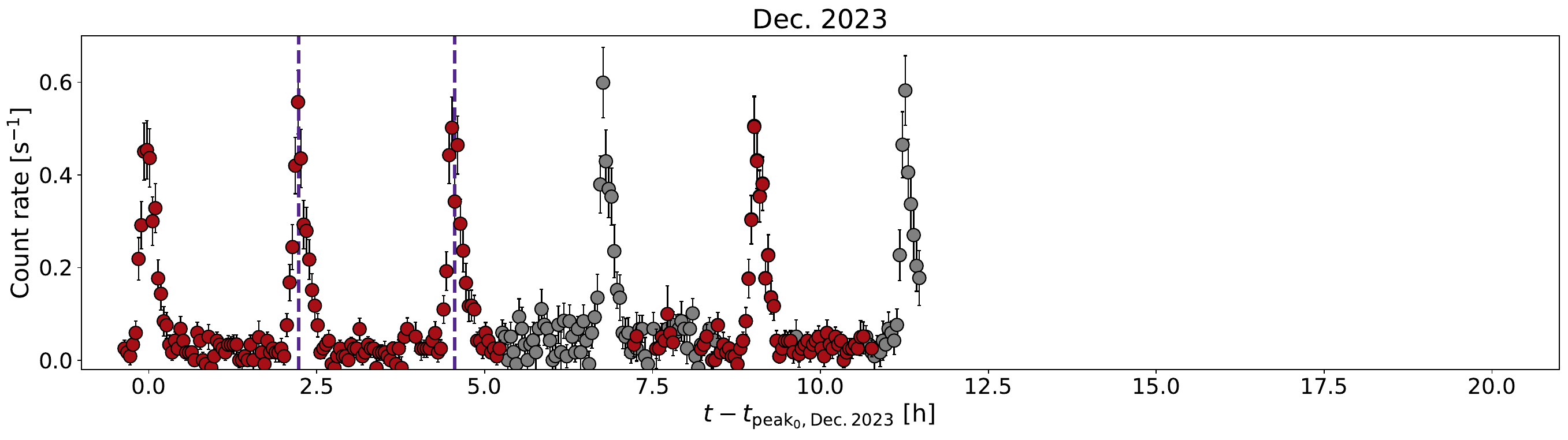}
		\caption{\emph{XMM-Newton} $0.2-2.0\,$keV light curves of the four epochs, scaled by the arrival time of the first eruption in each observation. Time intervals contaminated by the flaring background are shown in grey. In the top panel, we additionally show the rise, peak and decay phases selected for the Aug. 2020, as an example. In the second panel from the top, the first eruption is only partially resolved in EPIC-pn data and re-normalized MOS1 data are shown with empty symbols. As the x-axis range is shared, we note the shortening of recurrence times over the $\sim3.3$\,yr baseline, with vertical dashed lines showing the average arrival time interval of the previous epoch to guide the eye. 
        }
		\label{fig:lcus}
\end{figure*}

\begin{figure*}[tb]
		\centering
		\includegraphics[width=0.99\textwidth]{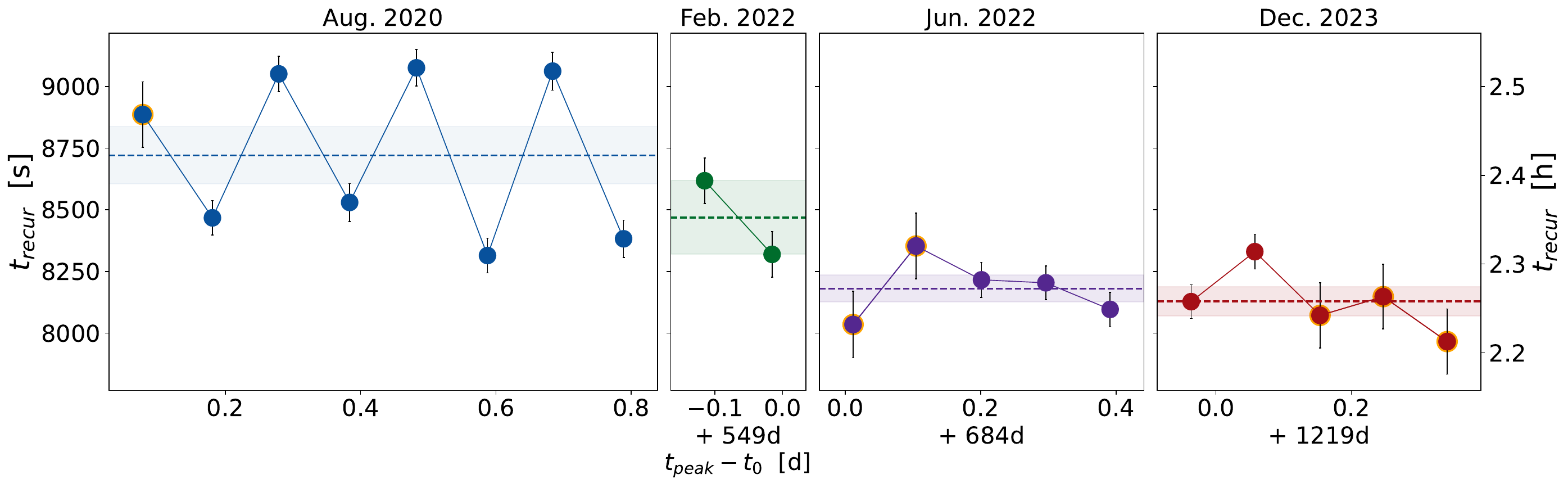}
		\caption{Evolution of the recurrence time between eruptions over the $\sim3.3\,$yr baseline. The start time $t_0$ represents the start of the Aug. 2020 observation. Data points with an orange contour contain an additional systematic uncertainty, as described in the text. The mean value in each epoch, with associated standard error of the mean, is shown with a dashed line and shaded contours.}
		\label{fig:period_evolution}
\end{figure*}

\begin{figure*}[tb]
		\centering
		\includegraphics[width=0.845\columnwidth]{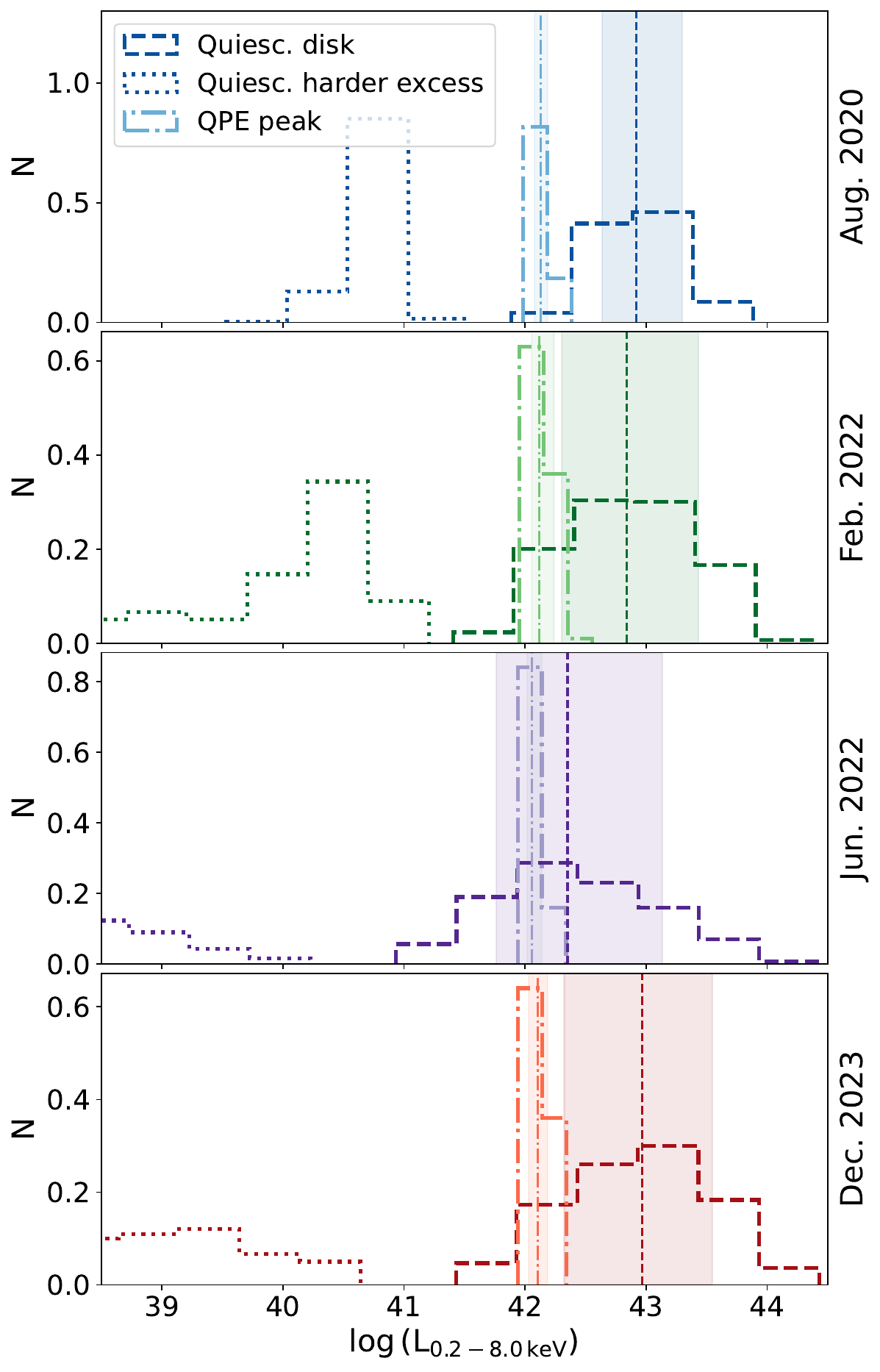}
        \quad
        \quad
        \includegraphics[width=0.845\columnwidth]{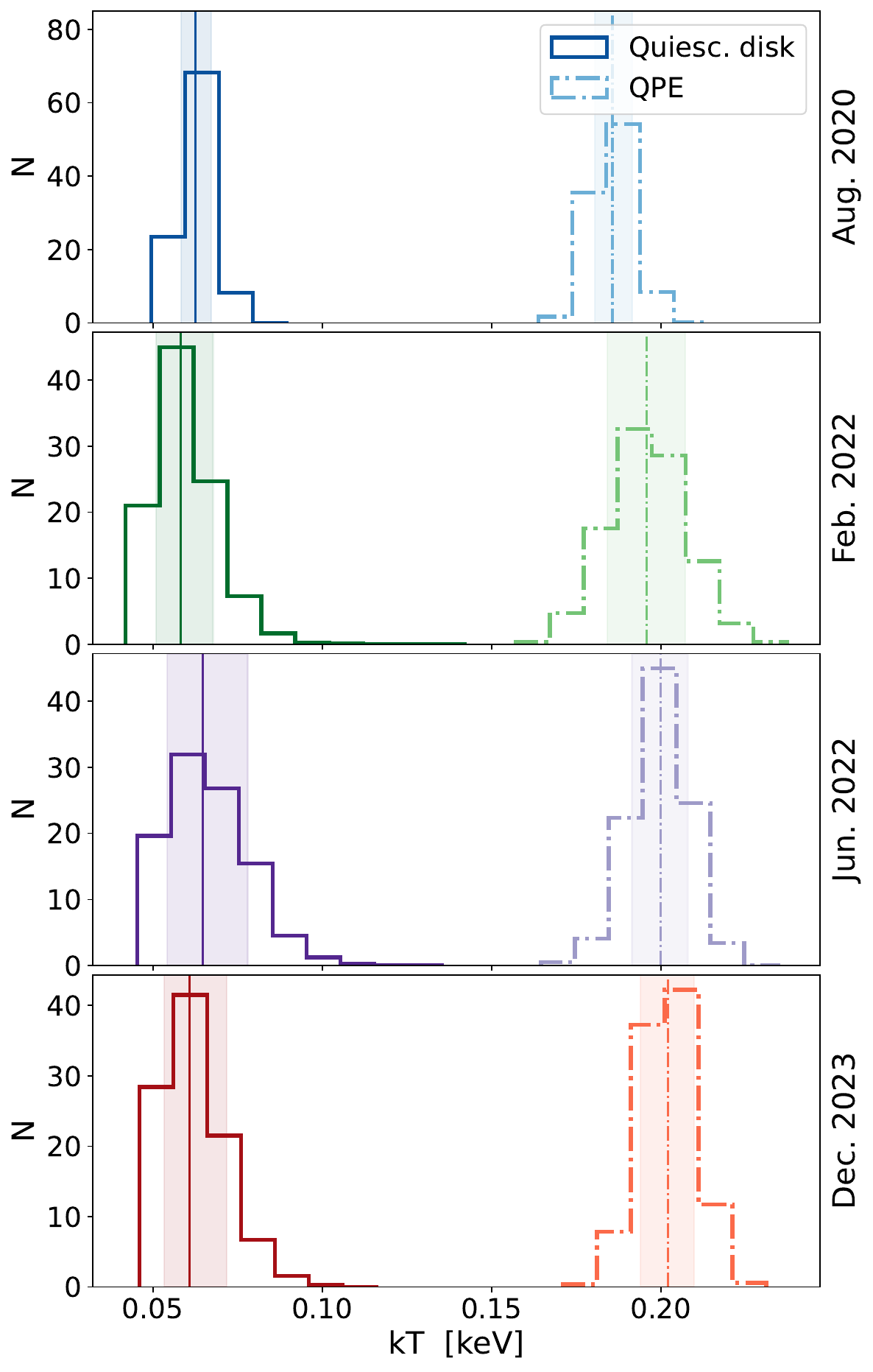}
		\caption{\emph{Left panels}: evolution of the $0.2-8.0\,$keV luminosity of individual spectral components (as labeled in the top panel for the first epoch). Histograms represent the fitted flux chains, converted to luminosity and corrected for absorption. Vertical lines and shaded area (following the same coding) highlight the median and 16th-84th interquintile range. There is no significant evolution of the accretion disk (dashed) nor the eruptions (dot-dashed), while the additional spectral component required for the quiescence spectrum in the first epoch becomes less significant and ultimately undetected at later epochs. 
        \emph{Right panels}: as the left panels, but showing posterior chains of the fitted temperature values. There is no significant evolution of the accretion disk temperature (solid) over the $\sim3.3\,$yr baseline. For the eruptions temperature (dot-dashed), despite the median increases over time and the first at last epochs are not compatible at $1\sigma$, all values are consistent within $2\sigma$.}
		\label{fig:flux_evolution}
\end{figure*}

In current datasets, the sparse light curve epochs of a given QPE source have gaps that are too long to maintain knowledge of both short-term and long-term evolution and, in particular, the number of eruptions occurred within the gaps, but some attempts have been made \citep[e.g.,][]{Xian+2021:qpemodel,Franchini+2023:qpemodel,Chakraborty+2024:ero1,Pasham2024ApJ...963L..47P,Zhou+2024:emri,Zhou+2024arXiv240506429Z}. In this work, we focus on the long-term ($\sim 3.3\,$yr) evolution of eRO-QPE2, which has thus far demonstrated regular flaring patterns and overall stable emission properties. We find that the QPE recurrence time decreases over time, albeit in a non-uniform way, while the long/short alternation seen in the first epoch \citep{Arcodia+2021:eroqpes} has not been significantly observed since. At the same time, the flux of both the eruptions and the quiescence emission has remained constant within uncertainties. We discuss possible mechanisms through which an EMRI system can reproduce these observations and the pitfalls. Our suggested scenario has clear predictions that future data will be able to verify.

\section{Observations and analysis}
\label{sec:res}

Here, we discuss the main results. We report the timing and spectral analysis in detail in Appendix~\ref{sec:processing}, using four \emph{XMM-Newton} datasets taken in Aug. 2020, Feb. 2022, Jun. 2022, and Dec. 2023, respectively. These are shown in Fig.~\ref{fig:lcus}.


\subsection{Decrease of the QPE recurrence time with constant flux}
\label{timingresult}

The recurrence time ($t_{\rm recur}$), defined as the peak-to-peak separation from light curve fits, was remarkable in the discovery epoch of Aug. 2020 in its alternating pattern between consecutively longer and shorter recurrences with difference at few percent level \citep{Arcodia+2021:eroqpes,Arcodia+2022:ero1_timing}. Similar alternating patterns were noted in GSN~069 and RXJ~1301.9+2747 \citep{Miniutti+2023:gsnrebr,Giustini+2020:qpe2}. We show this in the leftmost panel in Fig.~\ref{fig:period_evolution}, which reports the evolution of $t_{\rm recur}$. This alternating behavior has disappeared, within the uncertainties, over the rest of $\sim3.3\,$yr baseline probed by the data in this work (Fig.~\ref{fig:period_evolution}). At the same time, the mean recurrence time during a given epoch has decreased. The mean value (with associated standard error of the mean) decreased from $8721.0\,$s ($116.4\,$s) in Aug. 2020, to $8468.4\,$s ($149.3\,$s) in Feb. 2022, to $8180.8\,$s ($54.7\,$s) in Jun. 2022. After then, it only slightly decreased further to $8128.6\,$s ($59.5\,$s) in Dec. 2023, still compatible within uncertainties with the previous epoch. Using the first and last epoch, we infer an apparent $\Delta t_{\rm recur}/\Delta t \approx -6 \times 10^{-6}$\,s/s, or a $\sim 7\%$ decrease compared to the Aug. 2020 recurrence time. However, this decay was not uniform across all epochs and does not reproduce all of them. For instance, this recurrence decrease rate would overpredict the recurrence during Jun. 2022 by $\sim207\,$s, which is $\sim4$ times larger a difference than the Jun. 2022 uncertainty. As a matter of fact, there was a $\sim6\%$ decrease already between Aug. 2020 and Jun. 2022 ($\sim1.8\,$yr) and $\lesssim1\%$ (compatible with no decrease within uncertainties) in the following $\sim1.5\,$yr (see, e.g., Fig.~\ref{fig:QPE_Pdot_calc}).

Remarkably, all other properties of both flares and quiescent emission have remained constant within uncertainties and the available observations. In particular, the eruptions' duration remained consistent within the standard deviation of the values of each epoch: the mean and standard deviations are $1666 \pm 143\,$s, $1603 \pm 36\,$s, $1741 \pm 145\,$s, and $1736 \pm 151\,$s, for the different epochs, respectively. Here, we defined the rise-to-decay duration of eruptions from the difference between the times at which the Gaussian model adopted for the fit (Appendix~\ref{sec:processing}) reaches a value of $1/e^3$ compared to the peak. Similarly, the flux and temperature of both the eruptions and the quiescent components have remained constant within $3\sigma$ uncertainties. We show detailed fit values in Table~\ref{tab:ero2_spec} and the fit posteriors of all components in Fig.~\ref{fig:flux_evolution}, for flux and temperature in the left and right panels, respectively. It is perhaps noteworthy that the QPE peak temperature marginally (within $1\sigma$) increases from the first to the last epoch, although we stress the consistency of the measurements at $2\sigma$. Future data on much longer baselines will perhaps confirm or rule out this tentative trend. We computed the energy emitted by the eruptions (e.g., as in Giustini et al., subm.) as the integral of the Gaussian model used for the light curve fit, namely $\sqrt{2\pi}\, L_{\rm bol,QPE} \, \sigma_{\rm QPE}$, where $\sigma_{\rm QPE}$ is the best-fit Gaussian standard deviation (i.e., the characteristic timescale of the QPE duration). The resulting median (and related 16th-84th percentile) QPE energy is $1.2^{+0.2}_{-0.2}\times 10^{45}$\,erg, $1.0^{+0.3}_{-0.3}\times 10^{45}$\,erg, $1.1^{+0.2}_{-0.2}\times 10^{45}$\,erg, and $1.2^{+0.3}_{-0.2}\times 10^{45}$\,erg, respectively from Aug. 2020 to Dec. 2023. Thus, it is consistent across the epochs of our $\sim3.3\,$yr baseline.

Intriguingly, in the Aug. 2020 epoch an additional component, harder than the accretion disk, but still rather soft, is required to account for residuals (Appendix~\ref{sec:processing}). This component is only marginally constrained, albeit still statistically required, in Feb. 2020, but it is not statistically required after then (Fig.~\ref{fig:flux_evolution}). The nature of this component, occasionally seen in the quiescence of QPE sources \citep[e.g.,][]{Giustini+2020:qpe2,Chakraborty+2021:qpe5cand,Arcodia+2024:qpe34}, is still unclear. Here, we modeled it as Comptonization of accretion disk photons, but this interpretation is assumed and not driven by data. For this work, given the low signal-to-noise, we do not attempt further interpretation and defer to future work on all the QPE sources.

\subsection{The energy dependence of the eruptions}
\label{sec:edep}

\begin{figure}[tb]
		\centering
		\includegraphics[width=\columnwidth]{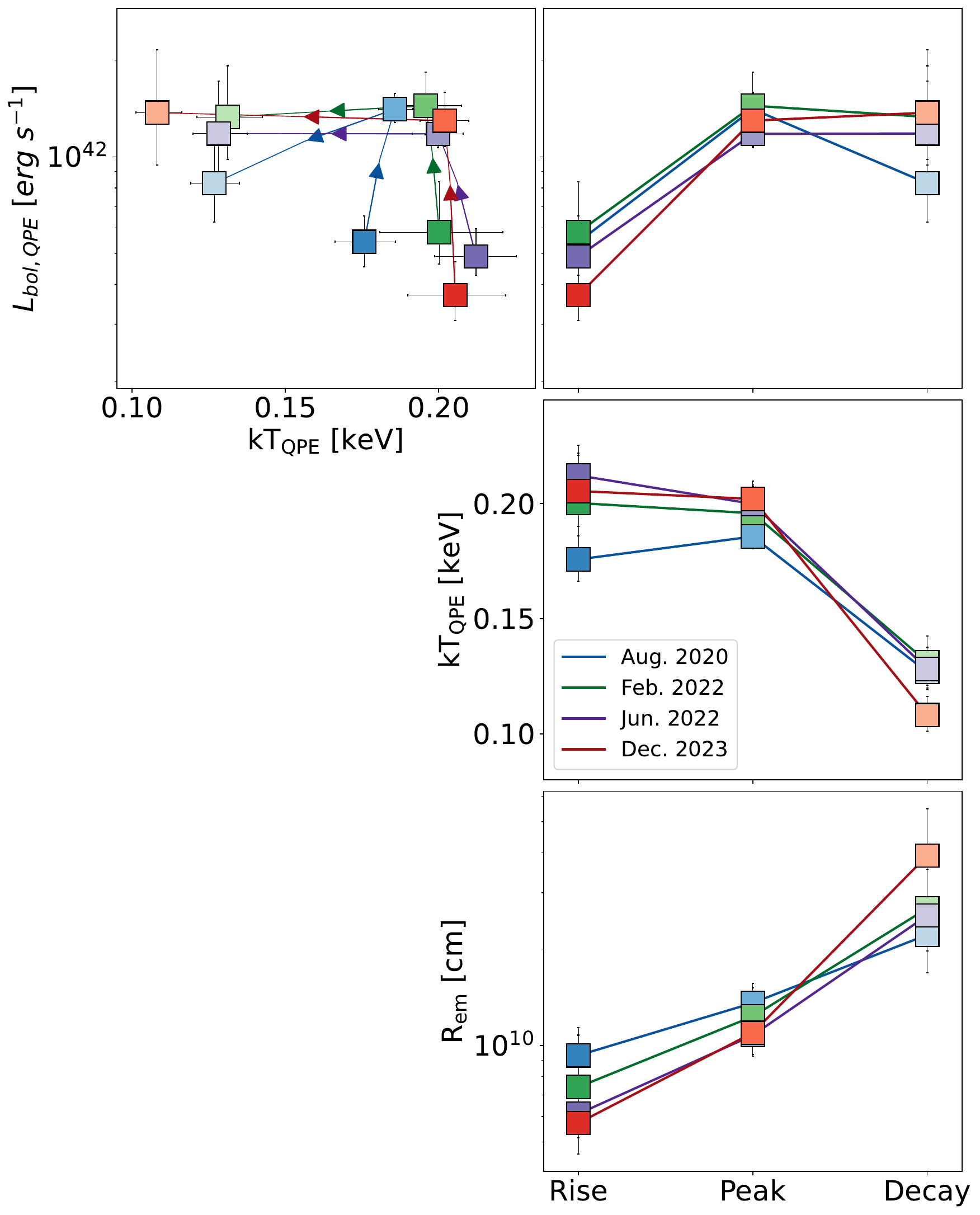}
		\caption{Spectral evolution of QPEs. Darker to lighter colors represent the evolution from start to end of the eruptions, color-coded as in the legend as a function of the epoch. The top-right (medium-right) panel shows the QPE bolometric luminosity (temperature) evolution with respect to the eruptions phase. The top-left panel shows their co-evolution. The bottom-right panel shows the evolution of the emitting radius, assuming a thermal spectrum and spherical geometry.}
		\label{fig:energy_evolution}
\end{figure}

\citet{Arcodia+2022:ero1_timing} reported the evolution of the spectral shape of eruptions in eRO-QPE1, which shows a harder spectrum during the rise than during the decay. Since then, this behavior has been found in the other QPE sources (\citealp{Miniutti+2023:gsnrebr}; Giustini et al., subm.; \citealp{Arcodia+2024:qpe34}), which is particularly useful to identify a common emission process among the growing number of repeating nuclear transients. Here, we test the same energy dependence in eRO-QPE2. Due to the lower signal-to-noise of eRO-QPE2 compared to other sources, we extracted a single rise and decay phase in addition to quiescence and peak, and we combine all eruptions within a single epoch. Fig.~\ref{fig:lcus} shows how the different phases were selected during the Aug. 2020 epoch, as an example, and Fig.~\ref{fig:spectra_Aug2020} shows the related spectra of the different phases. An analogous methodology was followed for the other epochs. After spectral analysis (Appendix~\ref{sec:processing}) we could isolate the eruption component and investigate the spectral evolution of the thermal component during the eruptions. Here, we confirm that the same spectral evolution is observed in eRO-QPE2 as well, consistently during each epoch (Fig.~\ref{fig:energy_evolution}). In the top-right (medium-right) panel of Fig.~\ref{fig:energy_evolution} we show the QPE bolometric luminosity $L_{\rm bol,QPE}$ (the temperature $kT_{\rm QPE}$) evolution with respect to the eruptions phase. We note that, comparably to the other known QPE sources, peak temperature is reached before peak luminosity, after which the spectrum cools down and it is the coolest during decay. The top-left panel of Fig.~\ref{fig:energy_evolution} shows the luminosity-temperature co-evolution during the various epochs. Darker to lighter colors represent the evolution from start to end of the eruptions, color-coded as in the legend as a function of the epoch. Indeed, the rise is always harder than the decay and, albeit with less data points, the hystheresis cycle is confirmed for eRO-QPE2 as well. Effectively, a thermal spectrum transiting from a hotter to a colder state at roughly comparable brightness could be attributed to an expansion of the associated black-body radius \citep{Miniutti+2023:gsnrebr,Chakraborty+2024:ero1}. In eRO-QPE2, we show this trend with time -- $R_{\rm em} \propto (L_{\rm bol,QPE}/T_{\rm QPE}^4)^{1/2}$ -- in the bottom-right panel of Fig.~\ref{fig:energy_evolution}. Intriguingly, despite the increase in $R_{\rm em}$ from rise to decay is seen consistently at all epochs, the rate of increase is larger at later epochs. In particular, the median rise-to-decay expansion of $R_{\rm em}$ is a factor $\sim2.4$, $\sim3.3$, $\sim4.3$, and $\sim7.0$, going from Aug. 2020 to Dec. 2023. Since the phase-folded light curve profiles appear consistent between Aug. 2020 and Dec. 2023 (Appendix~\ref{sec:processing} and Fig.~\ref{fig:phase_folded}), this is unlikely to be due to an evolution in the rise and decay phases over time, or biases due to how these phases are defined.



\section{Astrophysical interpretation}
\label{sec:disc}

The two main results from the long-term timing analysis of eRO-QPE2 are (1) the detection of a non-constant decrease in the QPE recurrence time $t_{\rm recur}$ and (2) an apparent disappearance of the long/short alternation pattern over the $\sim3.3\,$yr observational baseline (Fig.~\ref{fig:period_evolution}).

Here, we attempt to put these results into the context of orbital models. We note that orbital models have already been tested on eRO-QPE2. For instance, \citet{Franchini+2023:qpemodel} analysed the Aug. 2020 epoch of eRO-QPE2 as a test case for their model and \citet{Zhou+2024arXiv240506429Z} tested the 2020 and 2022 epochs independently, and, similarly to \citet{Linial+2023:qpemodel2}, suggested gas drag as the energy loss mechanism driving the apparent period decrease. However, the Dec. 2023 epoch (which highlights that a constant period decrease cannot reproduce the data, Fig.~\ref{fig:period_evolution}) was not included in their work, thus their conclusions do not grasp the full behavior shown by eRO-QPE2. A further complication is that without model-dependent assumptions timing fits are only possible for individual epochs of current datasets, the main obstacle being the lack on knowledge of the number of eruptions that occurred within the gaps. This is why we are unable to distinguish whether Fig.~\ref{fig:period_evolution} represents an overall continuous decreasing trend or whether the recurrence time and long/short alternation appearance/disappearance have also significantly oscillated over time. Thus, we discuss both options in Sect.~\ref{sec:sampling} and~\ref{sec:trend}. First, we outline in Sect.~\ref{sec:eqs} the main model assumptions and equations used in this work. 



\subsection{Orbital model assumptions and main equations}
\label{sec:eqs}

The interpretation we aim to test is that QPEs are triggered by a secondary orbiter of much smaller mass with respect to the primary black hole, which has $M_{\bullet}\sim 10^5$ M$_{\odot}$ \citep{Wevers+2022:hosts}, thus we take $M_{\bullet,5}=M_{\bullet}/10^5$M$_{\odot}$ as reference. The secondary is on a low-eccentricity orbit and is considered to be a either a solar-mass star, à la \citet{Linial+2023:qpemodel2}, or a $\sim40-100$~M$_{\odot}$ black hole, à la \citet{Franchini+2023:qpemodel}. For simplicity, we will refer to the orbital periodicity $P_{\rm orb}$ and its possible decrease $\dot{P}_{\rm orb}$ to explain the behavior shown by the observed quantity $t_{\rm recur}$. Even assuming two collisions per orbit, we note that $P_{\rm orb}$ is not exactly $t_{\rm recur,n}+t_{\rm recur,n+1}$. It would require, for instance, an idealized system with a fixed disk, the absence of precession and that flares are produced promptly at disk passage. Moreover, we will refer to orbital eccentricity $e_{\rm orb}$ and its possible decrease $\dot{e}_{\rm orb}$, to explain the presence first (Aug. 2020), and subsequent disappearance (Jun. 2022 and Dec. 2023), of the long/short alternation. However, the observed alternation actually probes a quantity which is an approximation of $e_{\rm orb}$, namely $e \equiv \delta t_{\rm recur}/\left< t_{\rm recur} \right>$, the difference between two consecutive recurrence times with respect to the average. For the scope of this work, we do approximate that trends in $t_{\rm recur}$ reflect trends in $\sim 1/2 P_{\rm orb}$ and that $e$ traces $e_{\rm orb}$. Hence, with this in mind Fig.~\ref{fig:period_evolution} unveils an apparent decrease for both $P_{\rm orb}$ and $e_{\rm orb}$.

\subsubsection{The relevant timescales}

The relevant timescales at play are the apsidal precession time of the secondary:
\begin{equation}
\label{eq:apsidal}
    \tau_{\star,\rm ap}\approx \frac{P_{\rm orb}}{3} \pfrac{r_0}{R_{\rm g}} \approx 22 \, {\rm d}\; \pfrac{\left< t_{\rm recur} \right>}{2.5 \, \rm hr}^{5/3} M_{\bullet,5}^{-2/3} \,,
\end{equation}
and, for non-zero spin of the primary black hole, the nodal precession timescale due to the Lense-Thirring \citep{Lense1918} effect of the spinning black hole on a misaligned orbit:
\begin{equation}
\label{eq:nodalstar}
    \tau_{\rm \star,LT} \approx \frac{1}{2 a_\bullet} P_{\rm orb} \pfrac{r_0}{R_{\rm g}}^{3/2} \approx
    1.6 \, {\rm yr} \; \pfrac{\left< t_{\rm recur} \right>}{2.5 \, \rm hr}^2 M_{\bullet,5}^{-1} a_\bullet^{-1} \,,
\end{equation}
where $a_\bullet$ is the SMBH's dimensionless spin, $P_{\rm orb} = 2  \left< t_{\rm recur} \right> $ is an approximation of the orbital period; $r_0$ is the orbit's semi-major axis,
\begin{equation}
    r_0 \approx 4.8 \times 10^{12} \, {\rm cm} \pfrac{\left< t_{\rm recur} \right>}{2.5 \, \rm hr}^{2/3} M_{\bullet,5}^{1/3} \approx
    323 \, R_{\rm g} \pfrac{\left< t_{\rm recur} \right>}{2.5 \, \rm hr}^{2/3} M_{\bullet,5}^{-2/3} \,,
\end{equation}
and $R_{\rm g}=GM_\bullet/c^2$ is the gravitational radius. An additional timescale could arise due to disk precession, although its effect depends on the exact model assumptions about its radial extent and mass distribution. \cite{Franchini2016} have shown that a compact, radiation pressure dominated accretion disk can rigidly precess around a spinning black hole on timescales as short as days, thus shorter than the apsidal precession of the companion. For instance, in \citet{Franchini+2023:qpemodel} the compact disk (e.g. as a result of a TDE) around a primary with spin $a_\bullet=0.5$ is misaligned by $\iota_{\rm d}=10^{\circ}$ with respect to the $x-y$ plane and precesses rigidly on a period of $\sim 5.7$ days. However, other more complex geometries may exist \citep[e.g.,][]{raj21,ivanov24}, for instance with warps or tilts around the collision radii, which would complicate the dynamics even further. Motivated by the relative regularity of QPEs in eRO-QPE2, we adopt a simple scaling between the nodal precession of the secondary ($\tau_{\rm \star,LT}$, Eq.~\ref{eq:nodalstar}) and that of the disk as $\tau_{\rm d,LT} = \alpha_{\rm LT} \tau_{\rm \star,LT}$, where $\alpha_{\rm LT}$ is a proportionality constant.

\subsubsection{Mechanisms for orbital energy loss}

For a black hole, neglecting the interaction with the disk, a low-eccentricity orbit inspiralling due to emission of gravitational waves (GWs) would yield a period derivative of order
\begin{multline}
    \left. \dot{P}_{\rm QPE} \right|_{\rm GW} \approx -\frac{96 \pi^{8/3}}{5} \frac{m_\star}{M_{\bullet}} (GM_\bullet/\left< t_{\rm recur} \right>)^{5/3} c^{-5} \approx \\
    -3\times 10^{-8} \; \pfrac{\left< t_{\rm recur} \right>}{2.5 \, \rm hr}^{-5/3} \pfrac{m_\star}{100 \rm M_\odot} M_{\bullet,5}^{2/3} \,.
\end{multline}
For a fixed period (or semi-major axis), an eccentric orbit would undergo more rapid inspiral ($\dot{P}_{\rm GW} \propto (1-e)^{-7/2})$). However, the relatively mild eccentricity inferred, $e \approx 0.06$ at most during Aug. 2020, would only increase $\dot{P}$ by less than a factor of 2 relative to a circular orbit. 
For the case of a star, $\dot{P}_{\rm QPE}$ is linearly lower with the mass. 

For a star, energy losses due to hydrodynamic gas drag are more relevant \citep[e.g.,][]{Linial+2023:qpemodel2,Linial+2024:14ko}. A star of mass $m_\star$ and radius $R_\star$, impacting a disk of surface density $\Sigma$ twice per orbit on a highly inclined, nearly circular orbit, would experience an orbital decay at an approximate rate
\begin{equation}
\label{eq:inclpdot}
    \left< \dot{P} \right> \approx - \frac{3\pi R_\star^2 \Sigma}{m_\star} \approx -2\times 10^{-6} \; \pfrac{m_\star}{\rm M_\odot}^{-1} \pfrac{R_\star}{\rm R_\odot}^{2} \pfrac{\Sigma}{10^5 \, \rm g \, cm^{-2}} \,.
\end{equation}
Naturally, the prediction from Eq.~\ref{eq:inclpdot} is sensitive to the largely unknown disk structure and surface density. 
Since the observed decrease in recurrence time was not uniform over the observed baseline (Fig.~\ref{fig:period_evolution}), disfavoring a constant $\dot{P}$ throughout the system's evolution, we further discuss the effect of inclination changes. As a toy model, we consider a star on a nearly circular orbit around a SMBH in the presence of a rigidly precessing accretion disk, both inclined with respect to the SMBH spin, with inclination angles $\iota_\star$ and $\iota_{\rm d}$, respectively. The relative inclination between the star and the disk thus varies periodically over the beat cycle, $\tau_{\rm LT,\star}/|1-\alpha_{\rm LT}|$, in the range $\iota_{\rm rel} \in (|\iota_{\star}-\iota_{\rm d}|$,$\iota_\star+\iota_{\rm d})$. At a given time, when the relative inclination between the star and the disk is $\iota_{\rm rel}$,
the orbital period decays at a rate
\begin{equation} \label{eq:Pdot_drag}
    \dot{P}(\iota_{\rm rel}) \approx -\frac{6\pi \Sigma R_\star^2}{m_\star} \frac{\sin^2{(\iota_{\rm rel}/2)}}{\sin{\iota_{\rm rel}}} \,.
\end{equation}
Here we assumed that the collision cross section is set by the star's physical size, and neglected any enhancement due to gravitational focusing of the impacted disk material (e.g., \citealt{Rein:2012,Arzamasskiy:2018})\footnote{This assumption is justified as long as $\iota_{\rm rel} \gtrsim \sqrt{(m_\star r_0)/(M_\bullet R_\star)} \approx 10^{-2}$, such that $Gm_\star/v_{\rm rel}^2 \lesssim R_\star$, where $v_{\rm rel} \approx 2 v_{\rm k} \sin{(\iota/2)}$ is the relative velocity between the star and the disk material upon impact. Eq.~\ref{eq:Pdot_drag} is also invalid for $\iota_{\rm rel} \lesssim h/r_0$, where $h$ is the disk's scale height, as the star is then entrained within the disk throughout most of its orbit, $\iota_{\rm rel} \lesssim h/r_0$, rendering the impulse approximation used here inappropriate.}. In the limit of small inclinations ($10^{-2} \lesssim \iota_{\rm rel} \ll 1$), $\dot{P} \propto \iota_{\rm rel}$. 
We further note that the orbital decay scales linearly with the dimensionless combination $\eta \equiv \Sigma R_\star^2/m_\star$, which we normalize to $\eta \equiv 10^{-6} \, \eta_{-6}$.

\subsubsection{Light travel time effects} 

Since the Aug. 2020 epoch showed significant long/short alternation, but following epochs did not, here we discuss the effect of light travel time delays. For instance, in \citet{Franchini+2023:qpemodel} these effects were included and considered small and negligible for most inclinations. However, the magnitude of this effect is maximal when the line connecting the two star-disk collision sites is near alignment with the observer's line of sight (requiring that both the disk and orbital plane are viewed near edge-on). Variations in recurrence time in this configuration for a circular stellar orbit, are of approximate magnitude
\begin{equation}
    e = \frac{\delta t_{\rm recur}}{\left< t_{\rm recur}\right>} \approx \frac{2r_0/c}{t_{\rm recur}} \approx 0.044 \; \pfrac{\left< t_{\rm recur} \right>}{2.5 \, \rm hr}^{-1/3} M_{\bullet,5}^{1/3} \,,
\end{equation}
similar to the observed $e \approx 0.05-0.06$ in August 2020. 

\subsection{On whether the behaviour shown in Fig.~\ref{fig:period_evolution} is dominated by random sampling of the orbital phase}
\label{sec:sampling}

Here, we discuss the scenario in which the evolution shown in Fig.~\ref{fig:period_evolution} for both $\dot{P}_{\rm orb}$ (decrease in $t_{\rm recurr}$) and $\dot{e}_{\rm orb}$ (decrease of alternation, or its disappearance) is dominated by modulations and random sampling of different orbital phases. These could be imprinted by changes in the relative inclinations between the orbit, the disk and the observer due to the precession frequencies at play, and light travel time effects, as shown in Sect.~\ref{sec:eqs}. 

We remind that this cannot be inferred by data in a model-independent fashion due to the gaps between observations. Thus, we discuss here all the possible impacts of these effects on the observations (Fig.~\ref{fig:period_evolution}). Even in case of a non-spinning primary, at least apsidal precession of the companion object would modulate the periodicity. That is, naturally, if the orbit is not circular to start with. For now, we assume that Aug. 2020 yields an apparent eccentricity close to the orbital one, thus of order of $\approx 0.05-0.06$, and the following epochs show a much lower variation ($\lesssim 0.01$) due to sampling effects. Considering the apsidal precession timescale on the order of $\approx20\,$d (Eq.~\ref{eq:apsidal}), alternation of $\lesssim 0.01$ are expected to occur for roughly $0.5 \, \rm d$, twice per precession cycle, corresponding to $\sim5\%$ probability if the source is randomly sampled (given by $2\times(0.01/0.06)/(2\pi) \approx 0.05$, the fraction of time where the orbital eccentricity vector is near alignment with plane of the disc). To have both Jun. 2022 and Dec. 2022 as a relatively rare aligned disk-orbit phase is even rarer ($\sim0.25\%$), but formally possible. In this case, a strong modulation would not change the average $P_{\rm orb}$ from epoch to epoch, thus orbital decay has to be invoked (see Sect.~\ref{sec:trend}).

If the primary black hole is spinning and the orbit and the disk are misaligned, both would evolve through nodal precession. As shown in detail in Sect.~\ref{sec:eqs}, we estimate the former as $\tau_{\rm \star,LT}\approx 1.6\,$yr and the latter as $\tau_{\rm d,LT} = \alpha_{\rm LT} \tau_{\rm \star,LT}$, where $\alpha_{\rm LT}$ is a proportionality constant smaller than one. The effect of $\tau_{\rm \star,LT}$, since it is much longer and comparable to the baseline, is discussed in Sect.~\ref{sec:trend}. Here, we discuss the possible sampling effect due to the shorter $\tau_{\rm d,LT}$. We first investigate whether disk alignment over the probed baseline of $\sim 3.3\,$yr could induce the observed decrease or disappearance of long/short alternation. Using the framework from \citet{Franchini+2023:qpemodel}, we find that the alignment process does not lead to a significant change in the recurrence pattern\footnote{We caution that we did not explore the disk alignment process self-consistently as we did not evolve the disk precession period with time. Therefore, the effect of possible changes in the disk structure (i.e. progressive alignment, differential precession) remains unexplored.}. We point out that the interplay between apsidal precession of the secondary and disk precession may give rise to more prolonged phases of low apparent eccentricity (compared to what we estimated above for when disk precession is absent), enhancing the probability of observing those phases by chance due to random sampling (Miniutti et al. in prep.). For instance, we evolve the solution for eRO-QPE2 from \citet{Franchini+2023:qpemodel} for $\sim 3$\,months, in order to get enough cycles, with the goal to estimate the likelihood of catching phases with low apparent eccentricity. Adopting $\tau_{\star,\rm ap}\sim 20\,d$, $\tau_{\rm d,LT}\sim 6$\,d, $P_{\rm orb}\sim5\,$h, and $e_{\rm orb}\sim0.05$ \citep{Franchini+2023:qpemodel}, we obtain that $\approx 12\%$ of the time the system is found at an apparent eccentricity $\lesssim 0.02$ and $\approx 6\%$ of the time lower than $\lesssim 0.01$. Similarly, but following opposite reasoning, the Aug. 2020 epoch could be the one when we caught the system in a rare phase of the orbit with high apparent eccentricity, which is otherwise low. Following the same calculation, we estimate that $\approx 18\%$ of the time the system is found at an apparent eccentricity $\gtrsim 0.10$. Thus, one can imagine that an intermediate $e_{\rm orb}$ between the values estimated for eRO-QPE2 could serendipitously give rise to samples with both higher and lower apparent eccentricity, such as the ones observed. In the same way, we estimate the impact of sampling this evolved model \citep{Franchini+2023:qpemodel} on the observed quantity $t_{\rm recur,n}+t_{\rm recur,n+1}$, which only seemingly traces the real $P_{\rm orb}$. Assuming $P_{\rm orb}=5\,$h, we note that the model predicts that $\sim25\%$ of the time the $t_{\rm recur,n}+t_{\rm recur,n+1}$ is found in the range of $5.0-5.4\,$h, and $\sim25\%$ of the time in the range of $4.9-5.0\,$h. Therefore, in principle sampling effect of a system such that in \citet{Franchini+2023:qpemodel} would be found at apparent periods significantly different than the true $P_{\rm orb}$. However, at least in the phases with a higher apparent period, consecutive recurrence times would appear to monotonically increase or decrease within a typical \emph{XMM-Newton} observation, which is instead not observed here. Furthermore, we note that if the disk is rigidly precessing \citep[e.g.,][]{Franchini+2023:qpemodel}, we should see modulation in the flux of the quiescence component, which is assumed to come from the innermost radii of the disk. However, we do not, as the quiescence flux remained remarkably constant over the probed $\sim3.3$\,yr (Fig.~\ref{fig:flux_evolution}). On the other hand, no significant flux change would be expected if the inner disk (where the quiescence comes from) is aligned and the outer (where the collisions happen) is not, or if there are large differences in how the inner and outer disk precess.

Finally, as the relative inclination between orbit, disk, and observer changes due to the precession of both disk and orbit, light travel effects would change. 
As we show in Sect.~\ref{sec:eqs}, the effect is in general small, although for edge-on inclinations and short observed periods like in eRO-QPE2 the effect could be non negligible ($\sim4\%$ of the QPE recurrence). Therefore, during the Aug. 2020 epoch we could have found the system in temporary edge-on alignment (i.e., light travel effects at few percent) and the following ones at higher inclinations (i.e., negligible light travel effects). As this evolution is expected to occur over $\sim \rm years$ timescale (comparable to $\tau_{\star, \rm LT}$), the observed trends from August 2020 to December 2023 appear in overall agreement with this proposed interpretation.

We conclude that any of these sampling effects are reasonable, but hard to estimate with the available data, thus any explanation would inherently be fine tuned at this stage. However, we note that these sampling effects are required to a larger extent if the orbiter is a black hole, compared to a star. As discussed in Sect.~\ref{sec:eqs} a star may lose more orbital energy than a black hole due to interactions with the disk, while for a BH the leading energy loss mechanism would be GW emission, much lower than the former and unable to explain current data alone. Quite crucially, this scenario in which sampling dominates the observed long-term evolution inherently predicts that we will likely have observations with a larger inferred recurrence time compared to Dec. 2023 (i.e. positive apparent $\dot{P}_{\rm orb}$) and/or with large long/short alternation again (i.e. positive apparent $\dot{e}$). This can be tested indirectly with future observations.

\subsection{On whether the behaviour shown in Fig.~\ref{fig:period_evolution} reflects a long-term trend}
\label{sec:trend}

Here, we discuss the alternative scenario in which the evolution shown in Fig.~\ref{fig:period_evolution} for both $\dot{P}_{\rm orb}$ (decrease in $t_{\rm recur}$) and $\dot{e}_{\rm orb}$ (decrease or disappearance of alternation) is dominated by a long-term trend. As much as some sampling effects due to precessions and inclination are expected (see Sect.~\ref{sec:sampling}) they would be second-order effects in this scenario. We observed a decrease in $P_{\rm orb}$ from $\sim4.85\,$h to $\sim4.53\,$h from Aug. 2020 to Dec. 2023 (summing two consecutive $t_{\rm recur}$ and averaging per epoch), corresponding to $r_0=320\,R_{g}$ and $r_0=307\,R_{g}$, respectively \citep[e.g.,][]{Franchini+2023:qpemodel}, and $e_{\rm orb}$ decreased from $\sim0.05-0.06$ to $\lesssim0.01$. It is easy to picture these two separate solutions separately, although ideally their evolution from one to the other can be modeled within the same framework. Taking the apparent period decrease from Aug. 2020 to Dec. 2023 at face value, we need a mechanism that induces $\dot{P}_{\rm orb} \approx -6\times 10^{-6}$. As shown in detail in Sect.~\ref{sec:eqs}, a BH orbiter would be disfavored since there is no mechanism to reproduce the observed decrease. Instead, for a star a modest orbital decay has been predicted to occur due to hydrodynamical drag, for instance as proposed in \cite{Linial+2023:qpemodel2} and \citet{Zhou+2024arXiv240506429Z}. Predictions do not exactly reproduce the observed value, but they are sensitive to the largely unknown disk structure and surface density (Sect.~\ref{sec:eqs}).

\begin{figure}
    \centering
    \includegraphics[width=\columnwidth]{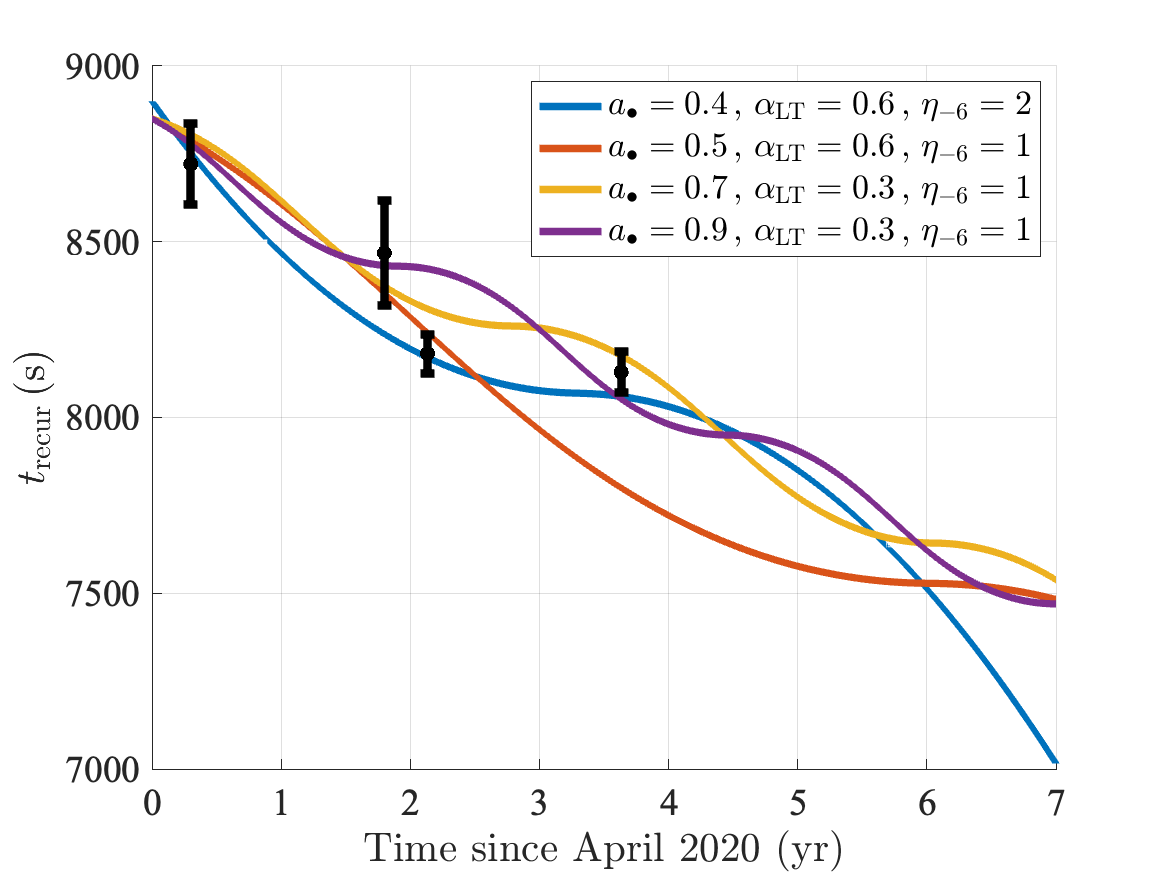}
    \caption{The observed recurrence time evolution, over-plotting some predicted evolution tracks due to gas drag and slowly evolving disk precession. Different lines are shown for differnet set of parameters (as shown in the legend), such as the spin ($a_\bullet$), the proportionality constant between disk and star nodal precessions ($\alpha_{\rm LT}$) and the parameter $\eta \equiv \Sigma R_\star^2/m_\star$. Variations in the slope of the calculated curves correspond to variations in the relative star-disk inclination, such that the plateaus (low $\dot{P}$) correspond to when the star and disk are near alignment.}
    \label{fig:QPE_Pdot_calc}
\end{figure}

What is instead harder to reproduce with this interpretation is the fact that the period decrease was non-constant. As a matter of fact, there was a $\sim6\%$ decrease between Aug. 2020 and Jun. 2022 and $\lesssim1\%$ in the following $\sim1.5\,$yr (Sect.~\ref{sec:res}, Fig.~\ref{fig:period_evolution}). In Sect.~\ref{sec:eqs}, we extended the hydro-drag interpretation with the effect of changing the relative inclination between the orbit and the disk due to $\tau_{\rm \star,LT}$ and $\tau_{\rm d,LT}$. In this framework, less orbital energy is lost when collisions are more face on, and viceversa. Now, we take Eq.~\ref{eq:inclpdot} and for a given set of parameters $\left\{ t_0 , P_0 , \eta , M_\bullet, a_\bullet, \iota_{\rm d} , \iota_{\star}, \alpha_{\rm LT} \right\}$, we calculate the orbital period $P(t) = P_0 + \int_{t_0}^t \dot{P}(t') \, dt'$. Fig.~\ref{fig:QPE_Pdot_calc} shows possible recurrence time evolution trends for a few different sets of parameters, as described in the legend, alongside the mean recurrence times at the four observed epochs. We fixed $M_\bullet = 10^{5} \, \rm M_\odot$, $\iota_{\rm d} = 0.8$ and $\iota_\star = 0.78$ (both approximately $45^{\circ}$) and manually adjusted the precession phase and $P_0$ to roughly align with the observations. Variations in the (negative) slope of the calculated curves correspond to variations in the relative inclination, such that the plateaus (low $\dot{P}$) correspond to when the star and disk are near alignment. The duration of these plateaus (i.e., the time spent around the minimal $\iota_{\rm rel}$) is proportional to the beat cycle of the star and disk nodal precession, which scales $\propto a_\bullet^{-1} |1-\alpha_{\rm LT}|^{-1}$. All sets of parameters lead to a monotonic decay in orbital period, and predict shorter recurrence times in the future. The transition between phases of rapid and slow decline could thus be attributed to the combined nodal precession of the star and the disc, coupled with hydrodynamical drag acting upon the star as it crosses the disk. We highlight that we do not show a fit to the data in Fig.~\ref{fig:QPE_Pdot_calc}, but that we do qualitatively reproduce the presence of non-constant $\dot{P}$ within reasonable ranges of the parameters above. We also note that other effects are in place when inclination changes (Sect.~\ref{sec:sampling}).

Even if this inclination effect is significant to explain the non-uniform $\dot{P}$ evolution, the related putative rate of eccentricity evolution is inconsistent with the relatively modest period evolution, if both orbital decay and circularization are due to the same hydrodynamical drag. As a matter of fact, we would expect in the low $e$ limit that $\dot{e}/e \sim \dot{P}/P$ (e.g., Appendix~B of \citealt{Linial+2023:qpemodel2}). Hence, unless we are missing on a fundamental efficient circularization process, some sampling effects such as temporary alignment of (only) the major axis with the disk orbital plane are required to increase the apparent eccentricity in Aug. 2020 compared to a true low $e_{\rm orb}$, as discussed in Sect.~\ref{sec:sampling}. 

Finally, we note that a concrete prediction in case the apparent evolution of Fig.~\ref{fig:period_evolution} is mostly an intrinsic period decrease trend, is that the mean QPE recurrence time should continue to decrease. If this is due to the above-mentioned drag picture, the recurrence time will decrease following alternating steeper and flatter epochs, following the trend with the relative star-disk inclination as shown in Fig.~\ref{fig:QPE_Pdot_calc}. Moreover, if the orbit has indeed circularized (regardless, for now, of the mechanism through which this has happened), no alternation will ever reappear. Conversely, if the orbit has not circularized and it is only an apparent transient alignment effect, long/short alternation phases may continue to reappear (and disappear) in future observations.

\subsection{On the constant flux of quiescence and eruptions}

Another key finding of this study is the striking consistency between the flux of both the quiescence and eruptions components across different epochs (Fig.~\ref{fig:flux_evolution}). The lack of significant disk evolution over this period is seemingly puzzling, if the origin of the disk is the accretion flow that follows the TDE of another star in the same galactic nucleus \citep[e.g.,][]{Linial+2023:qpemodel2,Franchini+2023:qpemodel}, as the late time X-ray emission from TDEs typically falls off on years timescale \citep[e.g.,][]{Mummery+2020}. However, as was recently demonstrated by \cite{Linial+2024:coupled}, mass (and thermal energy) added to the disk through ablation of the star's outer layers could result in a steady-state, equilibrium evolution of the disc, with its accretion rate matching the star's stripping rate, even if the initial disk was seeded by a TDE. Considering a low-mass star of $m_\star = 0.5 \, \rm M_\odot$, $\left< t_{\rm recur} \right> =2.5 \, \rm hr$, $M_\bullet=10^5 \, \rm M_\odot$ and applying results from Section 3.2 of \cite{Linial+2024:coupled}, we find that the self-sustained disk should settle to a steady-state accretion rate of roughly $\sim 0.1 \, \dot{M}_{\rm Edd}$, a surface density of about $\Sigma_{\rm d} \approx 10^4 \, \rm g \, cm^{-2}$, with an expected lifetime of order $10^2 \, \rm yr$. Hence, the lack of flux evolution in eRO-QPE2 over only $\sim3.3\,$yr is not a concern for this interpretation.

Next, we address the lack of substantial evolution in the flare temperature and luminosity over the observed baseline. If indeed the stability of the quiescent emission implies that the disk conditions remained nearly constant, an emission model that invokes collisions with the disk as the origin of the flares \citep[e.g.,][]{Linial+2023:qpemodel2,Franchini+2023:qpemodel} would indeed conform with stable flare properties (i.e., the star encounters similar disk conditions at the different epochs). However, as variations in the relative star-disk inclination may unavoidably be invoked to explain the system's non-trivial timing behavior (Sect.~\ref{sec:trend} and Sect.~\ref{sec:eqs}), the flare emission properties should be fairly insensitive to relative inclination, providing an important constraint on QPE emission models. Specifically, collisions of both low and high inclinations should produce flares of similar properties.

Finally, in Sect.~\ref{sec:res} we obtained that the rate of increase in the black body radius from rise to decay of the eruptions is apparently larger going from Aug. 2020 to Dec. 2023 (i.e. from a factor $\sim2.4$ to $\sim7.0$; Fig.~\ref{fig:energy_evolution}). This might be expected if the ejecta are not spherical like the black body $R_{\rm em}$ calculation assumes, if the system has evolved through changes in the inclination between the orbit and disk (hence collision and ejecta vector) and the observer. However, testing this effect quantitatively is beyond the scope of this work and the capabilities of the current emission models. We also stress that $R_{\rm em}$ does not necessarily represent the physical size of the X-ray emitting gas producing the flares. Specifically, in the context of the emission model discussed in \cite{Linial+2023:qpemodel2}, where the flares are generated by an expanding optically thick, shock-heated disk material, $R_{\rm em}$ does not trace the actual size of the emitting ejecta cloud (Vurm, Linial \& Metzger, in prep.).

\subsection{On the detectability of QPEs with GW detectors}

QPEs have been so far identified only in the electromagnetic domain through strong X-ray flares. The conjecture of QPEs being caused by EMRI systems however opens the possibility of multimessenger observations of these sources. Indeed, EMRI systems are primary targets of low frequency gravitational wave interferometers, like the recently adopted LISA mission \citep{2024arXiv240207571C}, planned to come online in the mid-thirties and covering the mHz band, as well as other mission concepts targeting even lower frequencies, e.g. $\mu$Ares surveying the $\mu$Hz band \citep{2021ExA....51.1333S}. EMRI systems that could explain the detected QPEs are systems that generally are from hundreds to several thousands of years away from the coalescence. 
EMRI sources producing QPEs with period $\sim2-20$ hours emit GWs at frequencies between $10^{-5}$ and $10^{-4}$ Hz. This frequency range represents the lower end of LISA window, where the sensitivity of the detector has undergone significant degradation. Nevertheless, the proximity of QPE sources, yielding a strain that can reach the level of $\sim 10^{-18}-10^{-17}$ in the case of a BH companion ($\sim40-100$\,$\rm M_\odot$), can compensate the LISA sensitivity loss. We can therefore expect signal-to-noise ratios (SNR) of order unity in the most optimistic cases. 
The situation can firmly improve with a detector targeting $\mu$Hz frequencies, like $\mu$Ares. There, the SNR of QPE sources is expected to be much higher and of the order $\sim10-100$, which would make them potentially detectable even in the case a stellar companion, depending on the exact mass of the central black hole and distance to the source.

\section{Summary}
\label{sec:concl}

In this work, we reported the long-term evolution of the QPE source eRO-QPE2 using four epochs over the $3.3\,$yr baseline since its discovery \citep{Arcodia+2021:eroqpes}. The main observational results are:
\begin{itemize}
    \item The average QPE recurrence time per epoch has decreased over time, albeit not uniformly: there was a $\sim 6\%$ decrease during the first $\sim1.8$\,yr and $\lesssim 1\%$ (compatible with no decrease within uncertainties) in the following $\sim1.5$\,yr (Fig.~\ref{fig:period_evolution} and Sect.~\ref{timingresult}).
    \item The long/short alternating behavior found at discovery has not been observed since. In particular, it was only minor $\sim0.8\,$yr after discovery (although only inferred with three consecutive eruptions) and consistent with being absent since then (Fig.~\ref{fig:period_evolution} and Sect.~\ref{timingresult}).
    \item The QPE duration has remained constant within uncertainties across all epochs (Sect.~\ref{timingresult}).
    \item The spectral properties, namely flux and temperature of both eruptions and quiescence components, have remained surprisingly consistent within $3\sigma$ uncertainties throughout all epochs (Fig.~\ref{fig:flux_evolution} and Sect.~\ref{timingresult}). The peak QPE temperature might have slightly increased going from the first to last epoch, although only at $1\sigma$.
    \item The energy dependence during eruptions follows the known QPE trend (\citealp{Arcodia+2022:ero1_timing,Miniutti+2023:gsnrebr,Arcodia+2024:qpe34}; Giustini et al., subm.), which shows a harder rise than decay. Modeling eruptions with a thermal spectrum, this leads to an increase in the fitted black body radius from rise to decay (Fig.~\ref{fig:energy_evolution}). The rate of this increase appears to have changed over time, towards larger increases at later epochs (Sect.~\ref{sec:edep}).
\end{itemize}

We attempted to interpret these results within orbital model prescriptions (Sect.~\ref{sec:eqs}), in particular the scenario of a secondary object (e.g., a star or a black hole with much smaller mass than the primary) on a low-eccentricity orbit which repeatedly pierces through the disk twice per orbit and produces the observed QPEs \citep[e.g.,][]{Linial+2023:qpemodel2,Franchini+2023:qpemodel}. In this framework, the QPE recurrence decrease is interpreted as (real or apparent) period decrease $\dot{P}$ and the decrease or disappearance of long/short alternation as (real or apparent) eccentricity decrease $\dot{e}$. However, since gaps between observations are too long we are not able to distinguish between an evolution dominated by just a decreasing trend, or by modulations (e.g. due to the orbit and disk precession frequencies at play). We therefore discuss these two scenarios separately (Sect.~\ref{sec:sampling} and~\ref{sec:trend}). If the current observations are dominated by unconstrained modulations no conclusive remarks are possible on the orbital evolution and the nature of the orbiter (Sect.~\ref{sec:sampling}). If instead the observed evolution (Fig.~\ref{fig:period_evolution}) is dominated by an intrinsic decreasing trend in $\dot{P}$ and $\dot{e}$ (Sect.~\ref{sec:trend}), observations can be qualitatively reproduced by a star losing orbital energy due to hydrodynamic gas drag from disk collisions. The time-dependent apparent $\dot{P}$ would be induced by a change in relative inclination between the orbit and the disk due to the orbiter's nodal precession, which for eRO-QPE2 is in the ballpark probed by our baseline (Fig.~\ref{fig:QPE_Pdot_calc}). However, the eccentricity decrease would be too fast to be only due to the same hydrodynamic gas drag regardless of the nature of the companion, thus some sampling effects (Sect.~\ref{sec:sampling}) would need to be invoked. If the behaviour shown in Fig.~\ref{fig:period_evolution} is indeed dominated by a trend, a black hole would not be able to lose enough orbital energy (through either GW emission or gas drag) to explain the observed decrease. A black hole orbiter interpretation does require the presence of modulations in both apparent period and eccentricity due to precession frequencies, which are however possible and remain untested with current data.

Fundamentally, these two cases (i.e., observation dominated by an intrinsic trend versus the effect of sampling different orbital phases, Sect.~\ref{sec:sampling} and~\ref{sec:trend}) come with very distinctive predictions for future X-ray data. In particular, in the case of a dominating decreasing trend we expect all future observations to show a shorter recurrence time than the latest epoch. Conversely, in the case of strong modulations dominating the observed evolution, we would expect at least some future observations to be found with a longer recurrence time, hence an apparent temporary period increase compared to the last epoch shown in this work. Future X-ray observations over the next months and years will be able to indirectly discern between these two scenarios. 

\begin{acknowledgements}
We thank the anonymous referee for their positive report. R.A. received support for this work by NASA through the NASA Einstein Fellowship grant No HF2-51499 awarded by the Space Telescope Science Institute, which is operated by the Association of Universities for Research in Astronomy, Inc., for NASA, under contract NAS5-26555. IL acknowledges support from a Rothschild Fellowship and The Gruber Foundation.
GM was supported by grant PID2020-115325GB-C31 funded by MICIN/AEI/10.13039/501100011033.  MG is supported by the ``Programa de Atracci\'on de Talento'' of the Comunidad de Madrid, grant numbers 2018-T1/TIC-11733 and 2022-5A/TIC-24235. MB acknowledges support provided by MUR under grant ``PNRR - Missione 4 Istruzione e Ricerca - Componente 2 Dalla Ricerca all'Impresa - Investimento 1.2 Finanziamento di progetti presentati da giovani ricercatori ID:SOE\_0163'' and by University of Milano-Bicocca under grant ``2022-NAZ-0482/B''. 
AS acknowledges the financial support provided under the European Union’s H2020 ERC Consolidator Grant ``Binary Massive Black Hole Astrophysics'' (B Massive, Grant Agreement: 818691).
GP acknowledges financial support from the European Research Council (ERC) under the European Union’s Horizon 2020 research and innovation program HotMilk (grant agreement No. 865637), support from Bando per il Finanziamento della Ricerca Fondamentale 2022 dell’Istituto Nazionale di Astrofisica (INAF): GO Large program and from the Framework per l’Attrazione e il Rafforzamento delle Eccellenze (FARE) per la ricerca in Italia (R20L5S39T9). This research benefited from interactions at the Kavli Institute for Theoretical Physics, during the KITP TDE24 program, supported by NSF PHY-2309135.
\end{acknowledgements}

%







\bibliographystyle{aa} 
\bibliography{bibliography} 



\begin{appendix}

\section{Data processing and analysis}
\label{sec:processing}

\begin{figure}[tb]
		\centering
		\includegraphics[width=0.99\columnwidth]{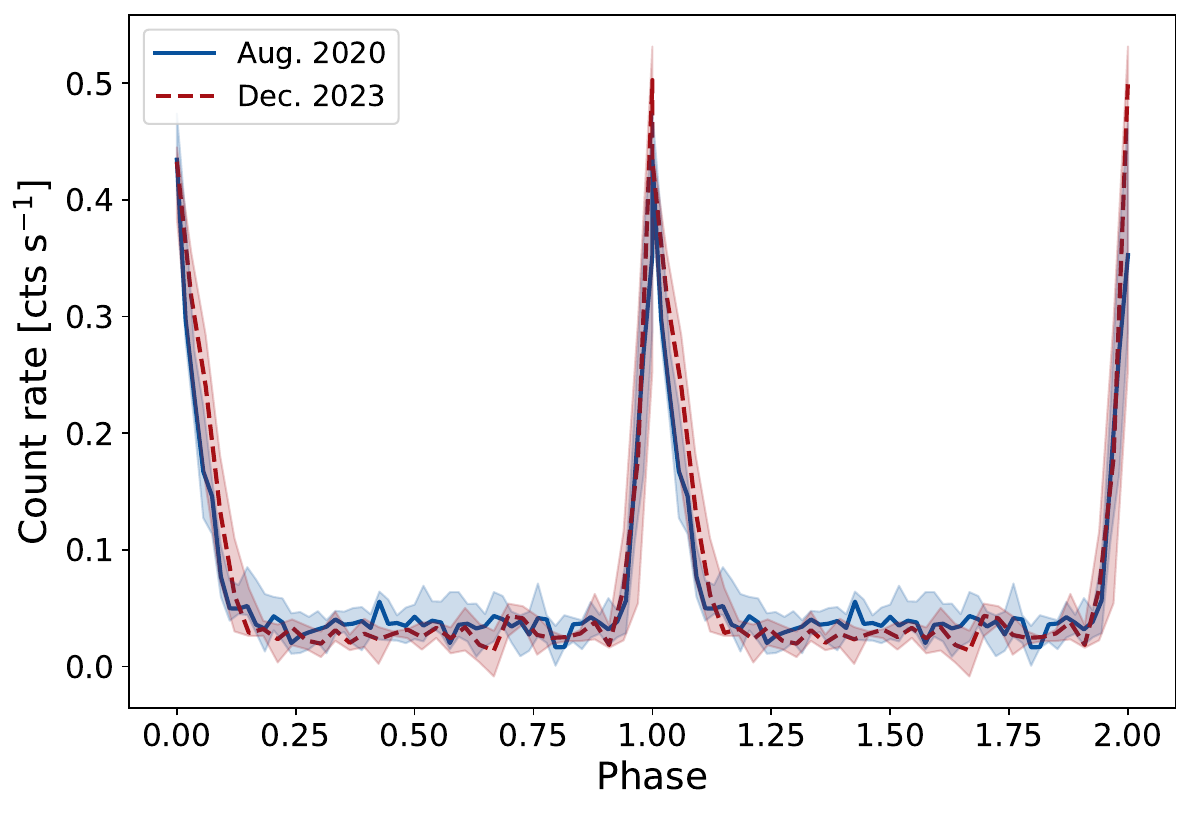}
		\caption{Median light curve profile (with the associated 16th and 84th percentile contours) for the Aug. 2020 (blue) and Dec. 2023 (red, dashed) epochs, folded at the eruption peaks.}
		\label{fig:phase_folded}
\end{figure}

Four proprietary or publicly available observations taken with \emph{XMM-Newton} were analyzed in this paper. The observations IDs are 0872390101, 0893810501, 0883770201, and 0931791301, and are referred to as `Aug. 2020', `Feb. 2022', `Jun. 2022', and `Dec. 2023', respectively, throughout this work. Data were reduced using SAS v. 20.0.0 and HEAsoft v. 6.31. Products were extracted from source (background) regions selecting a circle of $\sim30-40$" centered on eRO-QPE2 (in a nearby source-free region). For all epochs, we extracted and fitted light curves in the $\sim0.2-2.0\,$keV range (see Fig.~\ref{fig:lcus}). 
Timing analysis was performed using UltraNest \citep{Buchner2019:mlf, Buchner2021:ultranest} and assuming a Gaussian profile for the eruptions, which works generally well for eRO-QPE2 \citep{Arcodia+2021:eroqpes,Arcodia+2022:ero1_timing}. Asymmetric models \citep[e.g.,][]{Arcodia+2022:ero1_timing} will be systematically tested in future work, to account for the occasional residuals found during the decay in some eruptions. Here, we simply account for possible systematic uncertainties on the peak eruptions times by adding in quadrature the median difference between peak times obtained with the Gaussian and asymmetric models (which is $\sim57\,$s). For the Aug. 2020 dataset (top panel of Fig.~\ref{fig:lcus}), an additional eruption \citep[with respect to the light curve shown in the original discovery paper;][]{Arcodia+2021:eroqpes} was included in the EPIC-pn data by not performing background flaring screening. This is why the first data point of the Aug. 2020 dataset in Fig.~\ref{fig:period_evolution} is highlighted with an orange contour, and why it is shown in gray in the top panel of Fig.~\ref{fig:lcus}. For the Feb. 2022 dataset (second panel from the top in Fig.~\ref{fig:lcus}), the first eruption is only partial in the EPIC-pn dataset, therefore we used EPIC-MOS1 data to estimate the first recurrence time. Thus, we added in quadrature to the uncertainties on Feb. 2022 arrival times the average difference between EPIC-MOS1 and EPIC-pn arrival times from the full light curve fit ($\sim58$\,s). For the Jun. 2022 dataset (third panel form the top in Fig.~\ref{fig:lcus}), we extract the first three eruptions from background-contaminated intervals. For Dec. 2023, the same applies to the fourth and sixth eruption (bottom panel of Fig.~\ref{fig:lcus}). Thus, for these eruptions and recurrence times we proceeded in the same way as for the first eruption of Aug. 2020. All peak times shown with orange contours in Fig.~\ref{fig:period_evolution} (i.e., estimated in epochs contaminated by background flaring) have an additional error added in quadrature, estimated by the difference between arrival times or eruptions in light curves with or without background flaring filtering ($\sim 110\,$s). Similarly to \citet{Arcodia+2021:eroqpes}, we phase-fold light curve profiles at the eruption peaks for each epoch. In Fig.~\ref{fig:phase_folded} we show the median profile (with related 16th and 84th percentile contours) for the Aug. 2020 and Dec. 2023 epochs, in comparison. Despite a slight difference in the median profile towards longer or wider decays, the profiles are consistent at $1\sigma$ level.

\begin{figure}[tb]
		\centering
		\includegraphics[width=0.99\columnwidth]{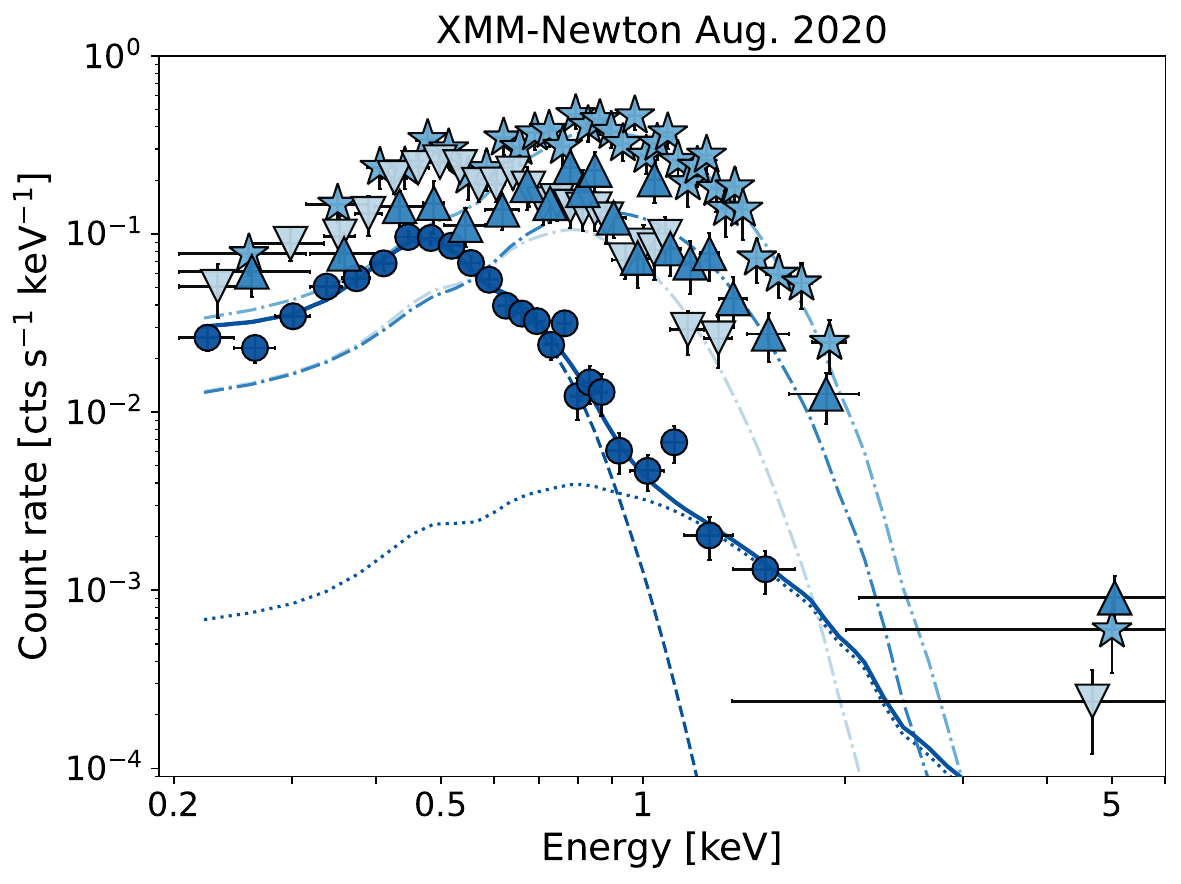}
		\caption{\emph{XMM-Newton} spectra for different phases, namely quiescence (circles), rise (up triangles), peak (stars), and decay (down triangles), extracted following the same colors and symbols as in the top panel of Fig.~\ref{fig:lcus}. The median models from the best-fit posterior are shown with a dashed (disk component, here \texttt{diskbb}), dotted (additional component, here \texttt{nthComp}) and dot-dashed (eruption component, here \texttt{zbbody}) line.}
		\label{fig:spectra_Aug2020}
\end{figure}

Event files from EPIC cameras were screened for flaring particle background for the purpose of X-ray spectral analysis. This was performed with the Bayesian X-ray Analysis software (BXA) version 4.0.7 \citep{Buchner+2014:BXA}, which connects the nested sampling algorithm UltraNest \citep{Buchner2019:mlf, Buchner2021:ultranest} with the fitting environment XSPEC version 12.13.0c \citep{Arnaud+1996:xspec}, in its Python version PyXspec\footnote{\href{https://heasarc.gsfc.nasa.gov/docs/xanadu/xspec/python/html/index.html}{Link to PyXspec}}. \emph{XMM-Newton} EPIC pn spectra were fit in the $0.2-8.0\,$keV band using \texttt{wstat}, namely XSPEC implementation of the Cash statistic \citep{Cash1979:cstat}. We adopted a Galactic column density of $N_H= 1.66 \times 10^{20} $~cm$^{-2}$ from HI4PI \citep{HI4PI+2016:HI4PI} and redshifted the source model to rest-frame using the spectroscopic redshift of $z=0.018$ \citep{Arcodia+2021:eroqpes}, corresponding to a luminosity distance of $\sim77\,$Mpc. Uncertainties are quoted using 16th and 84th percentiles ($\sim1\sigma$) vlues from fit posteriors, unless otherwise stated. The bolometric luminosity (labeled with ``bol'', e.g., $L_{bol,QPE}$ in Fig.~\ref{fig:energy_evolution} or $F_{bol}^{disk}$ in Table~\ref{tab:ero2_spec}) is obtained integrating the adopted source model between 0.001 and 100\,keV. For non-detections, we may quote $\sim1\sigma$ ($\sim3\sigma$) upper limits using the 84th (99th) percentiles of the fit posteriors. 

\begin{table*}[t]
	\footnotesize
	\setlength{\tabcolsep}{3pt}
	\caption{Spectral fit results for all epochs.}
	\label{tab:ero2_spec}
	\centering
	\begin{threeparttable}
		\begin{tabular}{cccccccccc}
			\toprule
            \multicolumn{1}{c}{Epoch} &
			\multicolumn{1}{c}{Spectrum} &
			\multicolumn{1}{c}{Model} &
			\multicolumn{1}{c}{$kT_{\rm disk}$} &
			\multicolumn{1}{c}{$F^{\rm disk}_{\rm 0.2-2.0\,keV}$} &
			\multicolumn{1}{c}{$F^{\rm disk,comp}_{\rm 0.2-2.0\,keV}$} &
			\multicolumn{1}{c}{$kT_{\rm QPE}$} &
			\multicolumn{1}{c}{$F^{\rm QPE}_{\rm 0.2-2.0\,keV}$} &
            \multicolumn{1}{c}{$F^{\rm disk}_{\rm bol}$} &
			\multicolumn{1}{c}{$F^{\rm QPE}_{\rm bol}$}
            \\
              &       &          & [eV]   &  [erg\,s$^{-1}$\,cm$^{-2}$]  & [erg\,s$^{-1}$\,cm$^{-2}$]  &  [eV]   &  [erg\,s$^{-1}$\,cm$^{-2}$] &  [erg\,s$^{-1}$\,cm$^{-2}$] &  [erg\,s$^{-1}$\,cm$^{-2}$]
            \\
			\midrule
             Aug. 2020 & Quiescence      &      \texttt{D+C}       &    $63\pm5$    &  $1.1^{+1.2}_{-0.6} \times 10^{-11}$  &    $7.5_{-2.7}^{+3.6} \times 10^{-14}$   &  -- &  --  &  $4.5^{+5.5}_{-2.5} \times 10^{-11}$   &  -- \\
                  & QPE rise    &     \texttt{D+C+BB}      &     $\sim63$   &  $\sim1.1 \times 10^{-11}$  &    $\sim7.5\times10^{-14}$   &  $176\pm10$   &  $7.3^{+1.4}_{-1.2} \times 10^{-13}$   &  -- &  -- \\
               & QPE peak    &     \texttt{D+C+BB}       &     $\sim63$   &  $\sim1.1 \times 10^{-11}$  &    $\sim7.5\times10^{-14}$   &  $186\pm6$   &  $(1.9\pm0.2) \times 10^{-12}$   &  -- &  $(2.0\pm0.2) \times 10^{-12}$ \\
               & QPE decay    &     \texttt{D+C+BB}       &     $\sim63$   &  $\sim1.1 \times 10^{-11}$  &    $\sim7.5\times10^{-14}$   &  $127\pm8$   &  $1.0^{+0.4}_{-0.2} \times 10^{-12}$   &  -- &  -- \\
             \midrule
             Feb. 2022 & Quiescence      &      \texttt{D+C}       &    $58_{-7}^{+10}$    &  $1.1^{+2.8}_{-0.8} \times 10^{-11}$  &    $<4.3 \times 10^{-14}$   &  -- &  --  &  $5.5^{+19.9}_{-4.5} \times 10^{-11}$   &  -- \\
                  & QPE rise    &     \texttt{D+C+BB}      &     $\sim58$   &  $\sim1.1 \times 10^{-11}$  &    $\lesssim 4.3 \times 10^{-14}$   &  $200\pm20$   &  $8.2^{+2.9}_{-1.6} \times 10^{-13}$   &  -- &  -- \\
               & QPE peak    &     \texttt{D+C+BB}       &     $\sim58$   &  $\sim1.1 \times 10^{-11}$  &    $\lesssim 4.3 \times 10^{-14}$   &  $196\pm11$   &  $1.9^{+0.6}_{-0.3} \times 10^{-12}$   &  -- &  $2.0^{+0.6}_{-0.3} \times 10^{-12}$ \\
               & QPE decay    &     \texttt{D+C+BB}       &     $\sim58$   &  $\sim1.1 \times 10^{-11}$  &    $\lesssim 4.3 \times 10^{-14}$   &  $131\pm11$   &  $1.6^{+0.8}_{-0.4} \times 10^{-12}$   &  -- &  -- \\
             \midrule
             Jun. 2022 & Quiescence      &      \texttt{D+C}       &    $65_{-11}^{+13}$    &  $3.7^{+22.5}_{-2.9} \times 10^{-12}$  &    $<3.7 \times 10^{-16}$   &  -- &  --  &  $1.8^{+12.2}_{-1.5} \times 10^{-11}$   &  -- \\
                  & QPE rise    &     \texttt{D+C+BB}      &     $\sim65$   &  $\sim3.7 \times 10^{-12}$  &    $\lesssim 3.7 \times 10^{-16}$   &  $212\pm13$   &  $6.6^{+1.5}_{-0.8} \times 10^{-13}$   &  -- &  -- \\
               & QPE peak    &     \texttt{D+C+BB}       &     $\sim65$   &  $\sim3.7 \times 10^{-12}$  &    $\lesssim 3.7 \times 10^{-16}$   &  $200\pm9$   &  $1.6^{+0.3}_{-0.1} \times 10^{-12}$   &  -- &  $1.7^{+0.3}_{-0.2} \times 10^{-12}$ \\
               & QPE decay    &     \texttt{D+C+BB}       &     $\sim65$   &  $\sim3.7 \times 10^{-12}$  &    $\lesssim 3.7 \times 10^{-16}$   &  $128\pm9$   &  $1.5^{+0.7}_{-0.4} \times 10^{-12}$   &  -- &  -- \\
            \midrule
            Dec. 2023 & Quiescence      &      \texttt{D+C}       &    $61_{-8}^{+10}$    &  $1.0^{+3.6}_{-0.8} \times 10^{-11}$  &    $<3.1 \times 10^{-15}$   &  -- &  --  &  $6.1^{+26.6}_{-5.1} \times 10^{-11}$   &  -- \\
                  & QPE rise    &     \texttt{D+C+BB}      &     $\sim61$   &  $\sim1.0 \times 10^{-11}$  &    $\lesssim 3.1 \times 10^{-15}$   &  $205\pm16$   &  $5.0^{+1.4}_{-0.9} \times 10^{-13}$   &  -- &  -- \\
               & QPE peak    &     \texttt{D+C+BB}       &     $\sim61$   &  $\sim1.0 \times 10^{-11}$  &    $\lesssim 3.1 \times 10^{-15}$   &  $202\pm8$   &  $1.8^{+0.3}_{-0.3} \times 10^{-12}$   &  -- &  $1.8^{+0.5}_{-0.3} \times 10^{-12}$ \\
               & QPE decay    &     \texttt{D+C+BB}       &     $\sim61$   &  $\sim1.0 \times 10^{-11}$  &    $\lesssim 3.1 \times 10^{-15}$   &  $108\pm8$   &  $1.7^{+0.8}_{-0.5} \times 10^{-12}$   &  -- &  -- \\
             \bottomrule
        \end{tabular}
	    \begin{tablenotes}
        \item Fit values show the median and related 16th-84th percentiles of the fit posteriors, using the disk (\texttt{D}), Comptonization (\texttt{C}) and black body QPE model (\texttt{BB}). Fluxes are reported in the rest-frame band and after correcting for Galactic and host absorption (see text for values). For the bolometric fluxes, the chosen energy range is $0.001-100\,$keV. For the `Aug. 2020' epoch, the different phases are shown in Fig.~\ref{fig:energy_evolution} and in the top panel of Fig.~\ref{fig:lcus}. Here, the quiescence spectrum, including the additional host galaxy absorption, is held fixed during the QPE epochs by letting its parameters free to vary only within the 10th-90th percentiles of the posteriors of the quiescence fit alone (hence the "$\sim$"). Given the spectroscopic redshift of $0.0175$ \citep{Arcodia+2021:eroqpes} and the cosmology adopted \citep{Hinshaw+2013:wmap9} the conversion for related luminosity values for eRO-QPE2 is $7.04\times10^{53}$\,cm$^2$.
        \end{tablenotes}
   \end{threeparttable}
\end{table*}

We extracted phase-folded spectra for quiescence, rise, peak, and decay selecting the related time intervals. We show this for the Aug. 2020 epoch in the top panel of Fig.~\ref{fig:lcus}, as an example, and follow a similar procedure for all epochs. We show the related spectra in Fig.~\ref{fig:spectra_Aug2020} for the Aug. 2020 epoch, with circles, up triangles, stars and down triangles for quiescence, rise, peak, and decay, respectively. Following the recent QPE literature, we model the quiescence as accretion disk emission (here \texttt{diskbb}) and possible residuals with an additional spectral component (here Comptonization with \texttt{nthComp}). Given the obvious presence of additional absorption \citep{Arcodia+2021:eroqpes}, we included in the fit a component at the redshift of the host galaxy (\texttt{ztbabs}): the median (and related 16th-84th percentiles) is $(0.47\pm0.07)\times 10^{22}\,$cm$^{-2}$, $(0.44\pm0.12)\times 10^{22}\,$cm$^{-2}$, $0.43^{+0.18}_{-0.14}\times 10^{22}\,$cm$^{-2}$, and $(0.51\pm0.14) \times 10^{22}\,$cm$^{-2}$, for epochs in chronological order, respectively. Thus, there is no apparent variability in absorption column across the epochs. We modeled the eruptions component with a thermal model (here \texttt{zbbody}). The quiescence spectrum, including the additional host galaxy absorption, is held fixed during the QPE epochs by letting its parameters free to vary only within the 10th-90th percentiles of the posteriors of the quiescence fit alone. We show the median models from the best-fit posterior of the Aug. 2020 epoch in Fig.~\ref{fig:spectra_Aug2020} with a dashed (disk component, here \texttt{diskbb}), dotted (additional component, here \texttt{nthComp}) and dot-dashed (eruption component, here \texttt{zbbody}) line. The unabsorbed luminosity posteriors for all epochs are shown in Fig.~\ref{fig:flux_evolution}, and the line style follows the same coding. We show in Table~\ref{tab:ero2_spec} the fit results and parameters for all epochs. We note that the additional harder component in quiescence is only statistically required for the Aug. 2020 at high significance, with an improvement in logarithmic Bayesian evidence of a factor $\Delta \log Z \sim1100$. However, while the flux chain of this additional component is only marginally constrained in Feb. 2022 (Fig.~\ref{fig:flux_evolution}), the fit significantly improves ($\Delta \log Z \sim64$). Thus, even if this component is not required in Jun. 2022 and Dec. 2023, we keep it in the best-fit model so that uncertainties on all parameters are marginalized over this potential spectral component. Finally, we note that we model this residuals as Comptonization given its ubiquity in accreting systems, although its nature in QPE sources is still up to debate. As further discussed in Sect.~\ref{sec:res}, the time evolution of eRO-QPE2 is remarkable because there is no significant evolution: the accretion disk and eruption component remain remarkably stable in flux and temperature over the course of the $\sim3.3\,$yr baseline (Table~\ref{tab:ero2_spec} and Fig.~\ref{fig:flux_evolution}).

\end{appendix}

\end{document}